\documentclass[11pt,journal,draftcls,onecolumn]{IEEEtran}
\usepackage{latexsym,amsfonts,amssymb,amsmath,bm}
\usepackage{algorithmic}
\usepackage{wrapfig}
\newcommand{\qed}{\hspace*{\fill}~\IEEEQED\par}

\newtheorem{Theorem}{Theorem}
\newtheorem{Theorem*}{Theorem}

\newtheorem{Claim*}{Claim}

\newtheorem{CounterExample*}{$\overline{\hbox{\bf Example}}$}
\newtheorem{Example}{Example}
\newtheorem{Example*}{Example}

\newtheorem{Intuition*}{Intuition}
\newtheorem{Joke*}{Joke}
\newtheorem{Lemma}[Theorem]{Lemma}
\newtheorem{Lemma*}{Lemma}
\newtheorem{Open problem}{Open problem}
\newtheorem{Proposition}{Proposition}

\newtheorem{Question*}{Question}
\newtheorem{Remark*}{Remark}
\newtheorem{Remark}{Remark}

\newtheorem{Lemmas*}{Lemma}

\def \bSubexa    {\begin{subexa}}

\newcommand{\ignore}[1]{{}}

\newcommand{\solution}[1]{}

\newcommand{\RR}{\mathbb{R}} 

\newcommand{\reals}{\RR}

\def \mynote#1{{}}

\def \Paren#1{{\left({#1}\right)}}

\def \Brack#1{{\left[{#1}\right]}}

\def\ignore#1{}

\newcommand{\bi}{\begin{itemize}}
\newcommand{\ei}{\end{itemize}}

\def\orpro{\mathop{\mathchoice
   {\vee\kern-.49em\raise.7ex\hbox{$\cdot$}\kern.4em}
   {\vee\kern-.45em\raise.63ex\hbox{$\cdot$}\kern.2em}
   {\vee\kern-.4em\raise.3ex\hbox{$\cdot$}\kern.1em}
   {\vee\kern-.35em\raise2.2ex\hbox{$\cdot$}\kern.1em}}\limits}

\def\andpro{\mathop{\mathchoice
 {\wedge\kern-.46em\lower.69ex\hbox{$\cdot$}\kern.3em}
 {\wedge\kern-.46em\lower.58ex\hbox{$\cdot$}\kern.25em}
 {\wedge\kern-.38em\lower.5ex\hbox{$\cdot$}\kern.1em}
 {\wedge\kern-.3em\lower.5ex\hbox{$\cdot$}\kern.1em}}\limits}

\def\circlei{\mathord{\mathchoice
 {\circle{4}\kern-.46em\lower.69ex\hbox{$i$}\kern.3em}
 {\circle{4}\kern-.46em\lower.58ex\hbox{$i$}\kern.25em}
 {\circle{4}\kern-.38em\lower.5ex\hbox{$i$}\kern.1em}
 {\circle{4}\kern-.3em\lower.5ex\hbox{$i$}\kern.1em}}\limits}

\renewcommand{\qed}{\Box}  

\newcommand{\ip}{X}

\newcommand{\re}{\hat{\ip}}
\newcommand{\loss}[1]{L_{#1}}

\newcommand{\losse}[1]{\hat{L}_{#1}}

\newcommand{\blde}{{\mathbf e}}

\begin{document}

\if@tmptwocolumn
\else
\renewenvironment{multline}{\begin{equation}}{\end{equation}}
\renewenvironment{multline*}{\[}{\]}
\fi

\title{Denoising as well as the best of \\ any two denoisers}
\author{Erik~Ordentlich%
\thanks{
E. Ordentlich (eordentlich@yahoo.com) is currently at Verizon Media. Work done while the author was with Hewlett-Packard Research Laboratories.  An abbreviated version of this submission appeared in the proceedings of the 2013 International Symposium on Information Theory, Istanbul, Turkey.}}
\date{\today}
\maketitle

\begin{abstract}
Given two arbitrary sequences of denoisers for block
lengths tending to infinity we ask if it is possible
to construct a third sequence of denoisers with an asymptotically
vanishing (in block length) excess expected loss relative to the best
expected loss of the two given denoisers for all clean channel input
sequences.  As in the setting of DUDE~\cite{Wei+05}, which solves this
problem when  
the given denoisers are sliding block denoisers, the construction is allowed to 
depend on the two given denoisers and the channel transition probabilities.
We show that under certain restrictions on the two given denoisers the
problem can be solved using a straightforward 
application of a known loss estimation paradigm.  We
then show by way of a
counter-example that the loss
estimation approach fails in the general case.  Finally, we
show that for the binary symmetric channel, combining the loss
estimation with a randomization 
step leads to a solution to the stated problem under {\em no}
restrictions on the given denoisers.
\end{abstract}


\section{Problem statement}
Given alphabets $ {\cal X} $ and $ {\cal Z} $,
an $n$-\emph{block denoiser} is a mapping $\re:{\cal Z}^n \rightarrow
{\cal X}^n$.
For any $z^n \in {\cal Z}^n$, let $\re(z^n)[i]$ denote the $i$-th term of
the sequence $\re(z^n)$.
Fixing a per symbol loss function $ \Lambda(\cdot,\cdot) $, for a
noiseless input sequence $x^n$ and the observed output  
sequence $z^n$, the \emph{normalized cumulative loss}
$\loss{\re}(x^n, z^n)$ of the denoiser $\re$ is
\[
\loss{\re}(x^n,z^n) = \frac{1}{n} \sum_{i=1}^n \Lambda\Paren{x_i,
  \re(z^n)[i]}.
\]
Given a
discrete memoryless channel (DMC) with transition probability matrix $ \Pi
$ between $ {\cal X}^n $ and $ 
{\cal Z}^n $ (i.e., the setting of DUDE~\cite{Wei+05}) and
two sequences of denoisers $ \hat{X}_1: {\cal Z}^n \rightarrow {\cal X}^n $
and $ \hat{X}_2: {\cal Z}^n \rightarrow {\cal X}^n $, we ask if there
always exists a sequence of denoisers $ \hat{X}_U $ whose expected
losses $ \loss{\hat{X}_U} $ satisfy
\begin{multline}
\limsup_{n\rightarrow \infty} \max_{x^n} E(\loss{\hat{X}_U}(x^n,Z^n))
\\
- \min\{E(\loss{\hat{X}_1}(x^n,Z^n)), 
E(\loss{\hat{X}_2}(x^n,Z^n))\} = 0.
\label{eq:univdef}
\end{multline}
Such a denoiser $ \hat{X}_U $ would then perform, in an expected
sense and asymptotically, as
well as the best of $ \hat{X}_1 $ and $ \hat{X}_2 $ for any
channel input
sequence(s).  

 The analogous problem in the settings of
prediction~\cite{Ces+97}, noisy prediction~\cite{WeiMer01}, and filtering (i.e.,
causal denoising)~\cite{Wei+07}
has been solved.  
DUDE~\cite{Wei+05} is a solution when the two
denoisers are sliding window denoisers (each denoised symbol is a
function of a window of noisy symbols centered at the corresponding
noisy symbol).  We are not aware of a solution to the problem at the
stated level generality.  In the sequel, we analyze the successes and
limitations of the loss estimator approach developed in~\cite{Wei+07}
for filtering and extended to the denoising setting
in~\cite{2dcxt}, in the context of the above
problem.  We show that while a direct application of
this approach fails in general, a 
certain randomized version
of the approach does, in fact, solve the above problem for the case of
the binary symmetric channel (BSC) (though, for now, not in a computationally
practical way).  The 
approach should be applicable to other DMCs, as will be addressed in
future work.

\section{Implications for error correction}
In a channel coding setting,
we can set the two target sequences of denoisers
to the decoders of {\em any} two sequences of channel codes with
vanishing maximal error probability.  A denoiser with
the universality property~(\ref{eq:univdef}) acts like 
a super-decoder that when applied to the decoding of the union of the
two sequences of codebooks achieves asymptotically vanishing {\em bit-error
rate} with respect to the transmitted codeword. It would
be interesting to know if such a super-decoder can be constructed without
relying on randomization, as we do herein.

\section{Loss estimator based approach}
\label{sec:lossestapproach}
A loss estimator for a denoiser $\re$ is a mapping 
$
\losse{\re}:{\cal Z}^n \rightarrow \reals
$
that, given a noisy sequence $z^n$, estimates the loss
$\loss{\re}(x^n,z^n)$ incurred by $\re$ to be
$\losse{\re}(z^n)$.  

Given a loss estimator, let $ \hat{j}^*(z^n) $ denote the index
$ j \in \{1,2\} $ of the denoiser $ \hat{X}_j $ attaining the
smallest estimated loss.  That is
$
\hat{j}^*(z^n) = \arg \min_{j \in \{1,2\}}
\losse{\hat{X}_j}(z^n).
$
Consider the loss estimator based denoiser
$
\hat{X}_U^n(z^n) = \hat{X}_{\hat{j}^*(z^n)}(z^n).
$

\begin{Lemma}
If for all $ \epsilon > 0 $, $ \losse{\hat{X}_j} $ satisfies
\begin{equation}
\limsup_{n\rightarrow \infty} 
\max_{x^n}\max_{j\in\{1,2\}} Pr(|\losse{\hat{X}_j}(Z^n) - 
\loss{\hat{X}_j}(x^n,Z^n)| \geq \epsilon)
 = 0
\label{eq:conc1}
\end{equation}
then $ \hat{X}_U $ satisfies~(\ref{eq:univdef}).
\label{lem:conc=univ}
\end{Lemma}

The proof of the lemma is similar to that of
Lemma~\ref{lem:conc=univrand} below, 
so we omit it.  
\if{false}
\noindent{\bf Proof.}
We'll give the proof for the first condition.  The second is similar.  
Let $ j^* $ denote
\[
j^*(x^n,z^n)= \arg \min_{j \in \{1,2\}}
\loss{\hat{X}_j}(x^n,z^n).
\]
Suppose $ x^n $ and $ z^n $ are such that 
$ |\losse{\hat{X}_j}(z^n) - 
\loss{\hat{X}_j}(x^n,z^n)| \leq \epsilon $ for $ j \in \{1,2\} $.
We then have
\begin{align*}
&\loss{\hat{X}_{\hat{j}^*}}(x^n,z^n) - 
\loss{\hat{X}_{j^*}}(x^n,z^n) \\
&= \loss{\hat{X}_{\hat{j}^*}}(x^n,z^n) -
 \losse{\hat{X}_{\hat{j}^*}}(z^n) +
\losse{\hat{X}_{\hat{j}^*}}(z^n) -
\loss{\hat{X}_{j^*}}(x^n,z^n) \\
&\leq \loss{\hat{X}_{\hat{j}^*}}(x^n,z^n) -
 \losse{\hat{X}_{\hat{j}^*}}(z^n) +
\losse{\hat{X}_{j^*}}(z^n) -
\loss{\hat{X}_{j^*}}(x^n,z^n) \\
& \leq 2\epsilon,
\end{align*}
implying that
\begin{multline}
Pr(
\loss{\hat{X}_{\hat{j}^*}}(x^n,Z^n) - 
\loss{\hat{X}_{j^*}}(x^n,Z^n) \geq 2\epsilon)
\\ \leq
\sum_{j=1}^2 
Pr(|\losse{\hat{X}_j}(Z^n) - \loss{\hat{X}_j}(x^n,Z^n)| 
\geq \epsilon).
\label{eq:probineq}
\end{multline}
Noting that $ \hat{X}_{\hat{j}^*} = 
\hat{X}_U $ it follows from this that
for all $ \epsilon > 0 $,
\begin{align}
\max_{x^n} & E(\loss{\hat{X}_U}(x^n,Z^n))
- \min\{E(\loss{\hat{X}_1}(x^n,Z^n)),
E(\loss{\hat{X}_2}(x^n,Z^n))\} \nonumber \\
& \leq 
\max_{x^n} E(\loss{\hat{X}_U}(x^n,Z^n)
- \loss{\hat{X}_{j^*}}(x^n,Z^n)) \nonumber \\
& \leq
2\epsilon + \Lambda_{\max}\max_{x^n}
Pr(
\loss{\hat{X}_U}(x^n,Z^n) -
\loss{\hat{X}_{j^*}}(x^n,Z^n) \geq 2\epsilon) \label{eq:foralleps}
\end{align}
The lemma now follows from~(\ref{eq:probineq}),
(\ref{eq:conc1}), and the fact that~(\ref{eq:foralleps}) holds for all
$ \epsilon > 0 $.
$ \qed $
\fi
\label{sec:losse}
Lemma~\ref{lem:conc=univ}
suggests that one solution to the problem of
asymptotically tracking the best of two denoisers is to estimate the
loss of each denoiser from the noisy sequence and denoise using the one
minimizing the estimated loss.  This would work provided the loss
estimator could be shown to satisfy~(\ref{eq:conc1}).

The following is
one potential estimator, first
proposed in~\cite{2dcxt}. 
The estimate of the loss incurred by {\em any} denoiser $\re$ proposed
in~\cite{2dcxt} is given by 
\begin{equation}
\label{eq:loss_estimate}
\losse{\re}(z^n) = \frac 1n \sum_{i=1}^n \sum_{x \in {\cal X}} h(x,z_i)
\sum_{z \in {\cal Z}} \Lambda(x, \hat{x}_i(z)) \Pi(x,z)
\end{equation}
where we use $\hat{x}_i(z)$ to abbreviate $\hat{X}(z_1^{i-1}\cdot z \cdot
z_{i+1}^n)[i]$ and $ h(\cdot,\cdot) $ satisfies
$
\sum_{z}\Pi(x,z)h(x',z) = 1(x = x').
$

\begin{Example}
For a DMC with invertible $ \Pi $,
$
h(x,z) = \Pi^{-T}(x, z),
$
uniquely.
\end{Example}
\begin{Example}
Specializing the previous example to a BSC with crossover probability
$ \delta $, 
\[
h(x,z) 
= \left\{
\begin{array}{ll}
\frac{\overline{\delta}}{1-2\delta} & (x,z) \in \{(0,0),(1,1)\} \\
\frac{-\delta}{1-2\delta} & (x,z) \in \{(0,1),(1,0)\}
\end{array}
\right.
\]
where $ \overline{x} $ defaults to $ 1-x $.
\end{Example}
\begin{Example}
For the binary erasure channel, $ h $ with the above property is not unique.
Consider a symmetric binary erasure channel with erasure probability $ 1/2 $.
One example of a valid $ h $ is:
\begin{equation}
h(x,z) = 2\cdot 1(x = z).
\label{eq:erasureh}
\end{equation}
In this case, the estimator~(\ref{eq:loss_estimate}) assumes an
especially intuitive form: 
 for each unerased symbol, determine what
the denoiser would have denoised that symbol to if had been erased,
 average the total losses over all symbols.  Formally,
\begin{equation}
\label{eq:erasure_loss_estimate}
\losse{\re}(z^n) = \frac{1}{n} \sum_{i:z_i \neq e} 
\Lambda(z_i, \hat{x}_i(e))
\end{equation}
and note that $ z_i = x_i $ for $ i: z_i \neq e $.
\end{Example}

\noindent{\bf Conditional unbiasedness.}  The loss
estimator~(\ref{eq:loss_estimate}) has been 
shown to be conditionally unbiased in the following sense.
Let 
\[
\tilde{\Lambda}_{i, \re}\Paren{z^n} \stackrel{\triangle}{=} \sum_{x
  \in {\cal X}} h(x, z_i) 
\sum_{z \in {\cal Z}} \Lambda(x, \hat{x}_i(z)) \Pi(x,z) 
\]
denote the estimate of the loss incurred on the $i$-th symbol.
Then
$
\losse{\re}(z^n) = \frac 1n \sum_{i=1}^n \tilde{\Lambda}_{i,\re}\Paren{z^n}.
$
\begin{Lemma}
\cite{2dcxt,MooWei09,Ord+13}
\label{unbiased}
For all $x^n$, all denoisers $\re$, and all $ i $, $1 \le i \le n$,
$z_1^{i-1}$, $z_{i+1}^n$ 
\begin{multline}
\label{eq:conditional_unbiased}
E\Brack{\tilde{\Lambda}_{i,\re}\Paren{Z^n} \left| Z_1^{i-1} = 
  z_1^{i-1}, Z_{i+1}^n = z_{i+1}^n \right.} \\ =
  E\Brack{\Lambda\Paren{x_i,\re(Z^n)[i]}\left| Z_1^{i-1} = 
  z_1^{i-1}, Z_{i+1}^n = z_{i+1}^n \right.} 
\end{multline}
and therefore
$ E\Brack{ \losse{\re}(z^n)} = E\Brack{\loss{\re}(x^n,Z^n)}. $
\end{Lemma}

\section{Success stories}
In this section, we review some special cases for which 
the loss estimator
(\ref{eq:loss_estimate}) 
 exhibits the concentration property (\ref{eq:conc1}) and hence for
 which the loss estimation paradigm solves the universal denoising problem.
A key tool is the martingale difference method for obtaining
concentration inequalities.  Briefly, in our context, consider a
function $ f:{\cal Z}^n \rightarrow \reals $ with $ E(f(Z^n)) = 0 $ and let 
\begin{equation}
M_i = E(f(Z^n)|Z^{i})
\label{eq:doobmartingale}
\end{equation}
denote the Doob martingale associated with $ f $ and $ Z^n $.
Let $ D_i = M_i - M_{i-1} $ and suppose it satisfies
$ |D_i| \leq c_i $ with probability one.   Then Azuma's inequality~\cite{mcdiarmid}
states that for any $ \epsilon > 0 $,
\[
Pr(f(Z^n) \geq n\epsilon) \leq e^{\frac{-n^2\epsilon^2}{2\sum_{i}c_i^2}}
\]
and
\[
Pr(f(Z^n) \leq -n\epsilon) \leq e^{\frac{-n^2\epsilon^2}{2\sum_{i}c_i^2}}
\]
from which it follows that
\[
Pr(|f(Z^n)| \geq n\epsilon) \leq 2e^{\frac{-n^2\epsilon^2}{2\sum_{i}c_i^2}}
\]

In our case, we will take $ f $ to be
\[
f(z^n) = \sum_{i} \tilde{\Lambda}_{i, \re}\Paren{z^n} 
- \Lambda(x_i,\hat{X}(z^n)[i]).
\]
We have $ E(f) = 0 $ by the unbiasedness of the loss estimator and 
noting that $ f $ is simply the difference between unnormalized
estimated and true losses, concentration inequalities for $ f $ are
precisely what we seek.

A special case of the above concentration inequalities is McDiarmid's
inequality~\cite{mcdiarmid} which applies to the case of $ Z_i $ being independent and
$ f $ satisfying
\[
|f(z_1^{i-1},x,z_{i+1}^n)-f(z_1^{i-1},y,z_{i+1}^n)| \leq c_i
\]
for all $ i $, $ x $, and $ y $.  This condition can be shown to imply
the above bound on $ D_i $ thereby yielding the above concentration
inequalities.  

Note that different concentration inequalities can be obtained by
conditioning $ f $ on $ Z_i $ in a different order than in
(\ref{eq:doobmartingale}), or even on increasingly refined functions
of $ Z^n $.  The best bound is obtained for which the resulting
martingale differences are the ``smallest''.

Finally, notice that the concentration inequality decays to zero even
if the $ c_i $ are as large as $ o(\sqrt{n}) $.  If even a single $
c_i = O(n) $, then no concentration is implied.

\begin{Example}
{\em Causal denoisers {\rm \cite{Wei+07}}.}
In this case, $ \hat{X}(z^n)[i] $ is a function of only $ z_1,\ldots,z_i
$.  It follows that
$
\Delta_i(z^n) = \tilde{\Lambda}_{i, \re}\Paren{z^n} 
- \Lambda(x_i,\hat{X}(z^n)[i])
$
is also causal,
 further implying, together with the conditional
unbiasedness, 
that $ D_i = \Delta_i(Z^n) $ in the above martingale difference
approach.  As this is clearly bounded by $ c \Lambda_{\max} $,
we have exponentially decaying concentration by the above inequality.
\end{Example}

\begin{Example}{\em Bounded (or slowly growing) lookahead denoisers.}
This is similar to the previous case, except that now $ D_i $ includes
the conditional expectations of a bounded number of additional terms.
The boundedness follows from the conditional unbiasedness and the
bounded lookahead.  The $ D_i $ are thus again bounded and exponential
concentration results.
\end{Example}

\begin{Example}{\em Each noisy sample affects only a few denoised values.}
If in a non-causal denoiser, the number of denoised values affected by
each noisy sample is $ 
o(\sqrt{n}) $, then concentration follows by McDiarmid's inequality above.
\end{Example}

\begin{Example}{\em Each denoised value depends only on a few noisy samples.}
\label{ex:sqrtnsamples}
Assume that for all $ i $, $ \hat{X}(z^n)[i] $ depends on
only $
c_n = o(\sqrt{n}) $ of the $ z_j $.  
For each $ i $ let $ V_i $ denote the
set of $ j $'s, such that $ \hat{X}(z^n)[i] $ depends on $ z_j $, and
for each $ j $ let $ S_j $ denote the
set of $ i $'s, such that $ \hat{X}(z^n)[i] $ depends on $ z_j $.
Let $ a_n $ satisfy $ a_n = o(\sqrt{n}) $ and $ c_n = o(a_n) $.
We then have
\begin{align*}
a_n|\{j:|S_j| > a_n\}| & \leq
\sum_j |S_j| \\ 
& = \sum_i |V_i| \\
& < nc_n
\end{align*}
so that
\begin{equation}
|\{j:|S_j| > a_n \}|  \leq \frac{c_n}{a_n} n.
\label{eq:Jbnd}
\end{equation}
Now note that
\[
Pr(|f(Z^n)| \geq n\epsilon) =
E(Pr(|f(Z^n)| \geq n\epsilon | \{Z_j:|S_j| > a_n\})) 
\]
so it suffices to show that the conditional deviation probabilities $
Pr(|f(Z^n)| \geq n\epsilon | \{Z_j:|S_j| > a_n\})) $ vanish with $ n $.
Let $ J = \{j:|S_j| > a_n\} $.
The idea is to note that
\[
f(Z^n) = \sum_{j \in J}\Delta_j + \sum_{j \notin J} \Delta_j
\]
and therefore that
\begin{align*}
Pr(|f(Z^n)| & \geq n\epsilon | \{Z_j:|S_j| > a_n\})
\\
& \leq 
Pr(\sum_{j \notin J}\Delta_j \geq n\epsilon-c|J| | \{Z_j:|S_j| >
a_n\}) \\
& \quad + 
Pr(\sum_{j \notin J}\Delta_j \leq -n\epsilon+c|J| | \{Z_j:|S_j| >
a_n\}).
\end{align*}
We can then apply McDiarmid's inequality conditionally to bound 
each of these conditional probabilities, since, 
the $ Z_j $ are independent, and
by design, for 
$ j \notin J $, each $
Z_j $ affects at most $ a_n = o(\sqrt{n})$ of the $ \Delta_j $, and
since the conditional expectation of $ \sum_{j\notin
  J}\Delta_j $ is 0 by the conditional unbiasedness of the loss estimator.
The
overall concentration follows from the fact that, by~(\ref{eq:Jbnd}),
$ |J| = o(n) $.

\end{Example}

The following proposition improves on this last example in terms of
expanding the number of noisy variables each denoising function can
depend on, but at the expense of non-exponential concentration.
\begin{Proposition}
Suppose for each $ i $, $ \hat{X}(z^n)[i] $ 
is a function of only (but
any) $
o(n) $ of the $ z^n $.  Then for any clean sequence $ x^n $
\begin{equation}
\max_{x^n} E([\losse{\hat{X}}(Z^n) - 
\loss{\hat{X}}(x^n,Z^n)]^2) = o(1)
\end{equation}
where the expectation is with respect to the noise.
\label{prop:functionoffew}
\end{Proposition}

\begin{Remark}
Note that, via an application of Chebyshev's inequality, 
this
proposition implies~(\ref{eq:conc1}).
\end{Remark}

\noindent{\bf Proof:}
Let $
\Delta_i = \Delta_i(z^n) = 
\tilde{\Lambda}_{i, \re}\Paren{z^n} 
- \Lambda(x_i,\hat{X}(z^n)[i])
$ 
so that
\begin{equation}
\losse{\hat{X}}(z^n) - 
\loss{\hat{X}}(x^n,z^n) = \frac{1}{n}\sum_{i=1}^n \Delta_i(z^n).
\label{eq:deltaequiv}
\end{equation}
Let $ T_i $
denote the subset of indices $ i $, such that $ \hat{X}(z^n)[i] $ is a
function of $ z_j $ with $ j \in T_i $.  We then have that $ \Delta_i
$ is a function of $ z_j $ with $ j \in T'_i = T_i \cup \{i\} $. 
We then have that for $ i $ and $ j \notin T'_i $,
\begin{align}
E(\Delta_i(Z^n)\Delta_j(Z^n)) &= 
E(E(\Delta_i(Z^n)\Delta_j(Z^n)|Z^{j-1},Z_{j+1}^n)) \nonumber \\
&= E(\Delta_i(Z^n)E(\Delta_j(Z^n)|Z^{j-1},Z_{j+1}^n)) \label{eq:zerocorra} \\
&= 0 \label{eq:zerocorr}
\end{align}
where~(\ref{eq:zerocorra}) follows from the fact that $ \Delta_i(Z^n)
$ is completely determined by $ Z^{j-1},Z_{j+1}^n $, since $ j \notin
T'_i $, and~(\ref{eq:zerocorr}) follows from the conditional
unbiasedness~(\ref{eq:conditional_unbiased}).

We then have
\begin{align}
E\big(\big(\sum_{i=1}^n \Delta_i\big)^2\big) & = 
\sum_{i=1}^n \sum_{j=1}^n
E(\Delta_i\Delta_j) \nonumber \\
& = \sum_{i=1}^n \sum_{j \in T'_i}
E(\Delta_i\Delta_j) \label{eq:zeroterms}\\
& = O(n\max_i|T'_i|)  = o(n^2), \label{eq:onsquared}
\end{align}
where~(\ref{eq:zeroterms}) follows from~(\ref{eq:zerocorr})
and~(\ref{eq:onsquared}) follows from the assumption of the
proposition. 
The proposition then follows after normalizing both sides
by $ n^2 $. 
$\qed$

\section{Problematic cases}
The following are some problematic cases for the above loss estimator based approach.

{\em Binary erasure channel with erasure probability 1/2.}
Consider the loss estimator based scheme with $ h(x,z) $
as given by~(\ref{eq:erasureh}) applied to tracking the two denoisers
\begin{align}
\hat{X}_1(z^n)[i] & = \sum_{j=1}^n 1(z_j = 0) \mod 2 \nonumber \\
\hat{X}_2(z^n)[i] & = 1+\sum_{j=1}^n 1(z_j = 0) \mod 2 \label{eq:paritydenoiser}
\end{align}
for each $ i $ that $ z_i = e $, under the Hamming loss.
Thus, denoiser 1 denoises to all 0's if the number of 0's in $ z^n
$ is even and to all 1's, otherwise, and denoiser 2 does precisely the
opposite.  Suppose the input sequence $ x^n $ is the all zero
sequence.  In this case (actually all cases), the expected (unnormalized)
loss of each denoiser 
is $ n/4 $.  It turns out, however, that the loss estimator based denoiser
always makes the worst possible choice.  Suppose $ z^n $ has an even
number of $ 0 $'s.  Denoiser 1 in this case achieves $0$ loss, while
denoiser 2 achieves a loss of $ N_e $ (denoting the number of erasures).  
Following~(\ref{eq:erasure_loss_estimate}), the estimated unnormalized loss
of denoiser 
1, on the other hand, is $
N_0
$
and of denoiser 2, $ 0 $.  The loss estimator based denoiser will thus
elect to follow denoiser 2, incurring a loss of $ N_e $.  The loss
estimator goes similarly astray for $ z^n $ with an odd number of $
0 $'s, and the average denoiser loss is thus $ n/2 $, failing to track
the $ n/4 $ average performance.

{\em Binary symmetric channel.}
\if{false}
Consider the adaptation of the denoiser pair (\ref{eq:paritydenoiser})
to the BSC with crossover $ \delta > 0 $, wherein the stated denoiser
output applies to all $ i $.  In this case, letting $ N_0 $ and $ N_1
$ respectively denote the number of $0$'s and $1$'s in $ x^n $, 
a straightforward analysis
of the loss estimator reveals that for $ z^n $ such that a denoiser
output's all 0's the estimated loss for that denoiser is $ N_0\delta +
N_1(1-\delta) $ while for the other set of $z^n$ the estimated loss is
$ N_0(1-\delta) + N_1\delta $.  If the input sequence is the
all-0 sequence we would expect $ N_0 $ and $ N_1 $ to be respectively
close to $ n(1-\delta) $ and $ n\delta $, implying that the estimator
based denoiser will select the denoiser outputting all 0's with high
probability (since $ 2\delta(1-\delta) < \delta^2 + (1-\delta)^2 $ for
all $ \delta \in (0,1) $).

Note, that although the loss estimator based denoiser is ultimately
operating correctly (and, in fact, substantially beating the $ n/2 $
average loss of either denoiser), the loss estimator fails to
concentrate around the true loss which is either $ n $ or $ 0 $.  It
seems just a happy accident that the loss estimator based denoiser
works in this case and indeed, we next specify a different parity
based denoising pair which breaks the loss estimator based denoiser on
the BSC.  
\fi
It turns out that the above example fails to break the
loss estimator based denoiser for the BSC and Hamming loss and a more
complicated example is required.
For the BSC with crossover probability $ \delta $, the
loss estimate of denoiser $ \hat{X} $ is
\begin{align}
&n \losse{\re}(z^n)  \nonumber \\
& {=}   \sum_{i:z_i = 0} \Big[
\frac{\overline{\delta}}{1{-}2\delta}(\delta \Lambda(0,\hat{X}(z^n{\oplus}
\blde_i)[i]){+}\overline{\delta}\Lambda(0,\hat{X}(z^n)[i]) \nonumber \\
& \; \quad {-}\frac{\delta}{1{-}2\delta}
(\delta
\Lambda(1,\hat{X}(z^n)[i]){+}\overline{\delta}\Lambda(1,\hat{X}(z^n{\oplus}
\blde_i)[i]) \Big] \nonumber \\
& \; {+} \sum_{i:z_i = 1} \Big[ \frac{\overline{\delta}}{1{-}2\delta}(\delta
  \Lambda(1,\hat{X}(z^n{\oplus} 
\blde_i)[i]){+}\overline{\delta}\Lambda(1,\hat{X}(z^n)[i]) \nonumber \\
& \; \quad {-}\frac{\delta}{1{-}2\delta}
(\delta
\Lambda(0,\hat{X}(z^n)[i]){+}\overline{\delta}\Lambda(0,\hat{X}(z^n{\oplus}
\blde_i)[i])  \Big],
\label{eq:bsclossest}
\end{align}
where $ \blde_i $ denotes the ``indicator'' sequence, with $
\blde_i[j] = 0 $ if 
$ j \neq i $ and $ \blde_i[i] = 1 $ and $ \oplus $ denotes
componentwise modulo two addition.
We can express~(\ref{eq:bsclossest}) in terms of the
joint type of the three sequences $ z^n, \hat{X}(z^n), $ and 
$ \{\hat{X}(z^n\oplus \blde_i )[i]\}_{i=1}^n $.  Specifically, for $
b_k \in \{0,1\} $, $ k = 0,1,2 $, define
\[
N_{b_0b_1b_2} = |\{i : 
z_i = b_0, \hat{X}(z^n)[i] = b_1, \hat{X}(z^n \oplus \blde_i)[i]=b_2
\}|,
\]
\[
N_{b_0b_1} = \sum_{b_2} N_{b_0b_1b_2}, \mbox{ and }
N_{b_0} = \sum_{b_1} N_{b_0b_1}.
\]

After some simplification, we can then express~(\ref{eq:bsclossest}) as
\begin{align}
n\losse{\re}(z^n) &=  -\frac{\delta}{1-2\delta}(N_{000}+N_{111}) +
\delta(N_{001} + N_{110}) \nonumber \\
& + \overline{\delta}(N_{010}+N_{101})
 + \frac{\overline{\delta}}{1-2\delta}(N_{011}+N_{100})
\label{eq:estlossNs}
\end{align}

For our example, we will set 
$
\hat{X}_1(z^n)[i] = \hat{X}_2(z^n)[i] = 0 
$
for all $ i $ and $ z^n $ with even parity.  Thus, for even parity,
the two denoisers will be identical, resulting in identical losses for
any clean sequence.  For $ z^n $ with odd parity, this implies that
the corresponding $ N_{b_0b_1b_2} = 0 $ for $ b_2 = 1 $ so that $
N_{b_0b_1} = N_{b_0b_10} $.  We will next
assume that the clean sequence is the all $ 0 $ sequence and specify
the behavior of the two denoisers for odd parity $ z^n $ taking this
into account.  Under this assumption on the clean sequence,
with probability tending to $ 1 $, $ 
N_1 = \delta n + o(n)$ and $ N_0 = 
\overline{\delta} n + o(n)$, so that for odd parity $ z^n $, with
probability tending to 1, we can write
\[
N_{10} = n\delta - N_{11} + o(n) \mbox{ and }
N_{00} = n\overline{\delta} - N_{01} + o(n).
\]
Using the above, 
 we can
further simplify~(\ref{eq:estlossNs}) to
\begin{align}
n\losse{\re}(z^n) &=  -\frac{\delta}{1-2\delta}N_{00} +
\delta N_{11} + \overline{\delta}N_{01}
+ \frac{\overline{\delta}}{1-2\delta}N_{10} \nonumber \\
&= N_{01}\left(\frac{\delta}{1-2\delta}+\overline{\delta}\right)
+ N_{11}\left(\delta -
\frac{\overline{\delta}}{1-2\delta}\right) + o(n) \nonumber \\
&=
(N_{01}-N_{11})\left(\frac{\delta}{1-2\delta}+\overline{\delta}\right)
+ o(n) .
\label{eq:estlossNssimp}
\end{align}

The two denoisers will then, respectively, denoise $ z^n $ with odd
parity 
so that:
\begin{align*}
\hat{X}_1(z^n) & \rightarrow N_{01} = 0, N_{11} = N_1 \\ 
\hat{X}_2(z^n) & \rightarrow N_{01} = \lfloor \delta N_0 \rfloor ,
N_{11} = 0.
\end{align*}
Thus, denoiser 1, for $ z^n $ with odd parity, sets $
\hat{X}_1(z^n)[i] = z_i $, while denoiser 2 sets
$ \hat{X}_2(z^n)[i] = 0 $ if $ z_i = 1 $ and 
$ \hat{X}_2(z^n)[i] = 1 $ for an arbitrary fraction $ \delta $ of
those $ i $ for which $ z_i = 0 $.   Under the assumption that $ x^n
$ is all $ 0 $, the following summarizes the actual losses and
estimated losses for $ z^n $ with odd parity and $ N_1 = n\delta +
o(n) $:
\[
\begin{array}{|l|l|l|}
\hline \mbox{Denoiser} & n\loss{\re}(x^n,z^n) & n\losse{\re}(z^n)
\\ 
\hline 1  &  \delta n + o(n)  &
-\left(\frac{\delta}{1-2\delta}+\overline{\delta}\right)\delta n +
o(n) \\
2  & \delta\overline{\delta}n + o(n) &
\left(\frac{\delta}{1-2\delta}+\overline{\delta}\right)\delta\overline{\delta}
n + o(n) \\
\hline
\end{array}
\label{eq:distvsestdist}
\]
Thus, we see that the estimated loss for denoiser 1 is smaller
(negative in fact) while its actual loss is larger.  Since the above
scenario ( odd parity $ z^n $ and $ N_1 = \delta n + o(n) $) occurs
roughly with probability $ 1/2 $, and for $ z^n $ with even parity the
two denoisers both incur zero loss, it follows that the expected
loss of the loss
estimator based denoiser fails to track the expected loss of the best denoiser,
namely denoiser 2, in this case.

\section{Smoothed denoisers}
The misbehavior of the loss estimator in the previous section appears
to be the result of an excessive sensitivity of the target
denoisers to the noisy sequence.  Our path forward for the BSC is to first
``smooth'' the target denoisers via a randomization procedure in a
way that does not significanly alter their  
average case performance on any sequence.  The expected performance
(with respect to the randomization) of
the smoothed denoisers, in
turn, will be shown to be more amenable to accurate loss estimation.  
To this end, for the BSC-$\delta$ case, let $ W^n $ be
i.i.d. Bernoulli-$q_n$ for some $ q_n $ vanishing (with $n$). Given a 
denoiser $ \hat{X} $, the randomized (smoothed) version is taken to be
\begin{equation}
\hat{X}'(z^n) = \hat{X}(z^n\oplus W^n).
\label{eq:randdendef}
\end{equation}
Conditioned on $ Z^n=z^n $, the expected loss (with respect to $ W^n
$) of this randomized denoiser
is
\if{false}
\footnote{For any quantity $ X $ depending on the randomization $
  W^n $, the notation $ \overline{X} $ will denote the
  expectation  of $ X $ with respect to the randomization $ W^n $,
  with all other variables fixed.  This should not be confused with
  the case of a real number $ x \in [0,1] $, for which
  $ \overline{x} = 1-x $, as above.  The intended meaning 
  of this notation should be clear from the context.}
\fi
\begin{equation}
\overline{L}_{\hat{X}'}(x^n,z^n) \stackrel{\triangle}{=}
\frac{1}{n}\sum_{i} E_{W^n}\Lambda(x_i,\hat{X}(z^n\oplus W^n)[i]) \label{eq:randomizedloss} .
\end{equation}
We can readily adapt the above loss estimator to estimate
$ \overline{L}_{\hat{X}'}(x^n,z^n) $ 
as
\begin{multline}
\hat{{\overline{L}}}_{\re'}(z^n)  = \frac 1n \sum_{i=1}^n \sum_{x \in {\cal X}} h(x,z_i) \\
\times \sum_{z \in {\cal Z}} E_{W^n}\Lambda(x,\hat{X}((z^{i-1},z,z_{i+1}^n)\oplus W^n))[i] \Pi(x,z).
\label{eq:loss_estimate_randomized}
\end{multline}
The summands (over $ i $) of this estimate of the expected loss of
the randomized 
denoiser also have
a conditional unbiasedness property.
Specifically,
letting
\begin{multline}
\overline{\tilde{\Lambda}}_{i,\hat{X}'}(z^n) \stackrel{\triangle}{=}
\sum_{x \in {\cal X}} h(x,z_i) \\ \times
\sum_{z \in {\cal Z}} E_{W^n}\Lambda(x,\hat{X}((z^{i-1},z,z_{i+1}^n)\oplus W^n))[i]
\Pi(x,z),
\label{eq:overlinetildelambda}
\end{multline}
we have
\begin{align}
&E\Big[\overline{\tilde{\Lambda}}_{i,\re'}\Paren{Z^n} \left| Z_1^{i-1} = 
  z_1^{i-1}, Z_{i+1}^n = z_{i+1}^n \right.\Big] \nonumber \\ &=
  E\Big[E_{W^n}\Lambda\Paren{x_i,\re(Z^n{\oplus} W^n)[i]}
\left| Z_1^{i-1}{=} 
  z_1^{i-1}, Z_{i+1}^n {=} z_{i+1}^n \right.\Big]
\label{eq:conditional_unbiased_rand}
\end{align}

We can then prove (see below) the following key lemma.
\begin{Lemma}
For all $ \delta $ and $ \hat{X'} $ as in~(\ref{eq:randdendef}) with $
q_n = n^{-\nu} $ and 
$ 0 < \nu < 1 $, 
\begin{equation}
\max_{x^n}
E\left(\overline{L}_{\hat{X}'}(x^n,Z^n)-\hat{\overline{L}}_{\hat{X}'}(Z^n)\right)^2
= o(1) 
\end{equation}
with $\overline{L}_{\hat{X}'} $ and $ \hat{\overline{L}}_{\hat{X}'} $ as
in~(\ref{eq:randomizedloss}) and~(\ref{eq:loss_estimate_randomized})
and where the expectation is with respect to the BSC-$\delta$ induced
$ Z^n $.
\label{lem:randden} 
\end{Lemma}

The lemma implies that for any BSC the 
estimate~(\ref{eq:loss_estimate_randomized}) of the 
randomized denoiser conditional expected loss concentrates for all
clean sequences and all underlying denoisers, including
those in which the
estimate of the underlying denoiser loss does not.   This motivates
an estimation minimizing randomized denoiser which departs
from the approach of Section~\ref{sec:lossestapproach} as follows.
Given denoisers $ \hat{X}_1 $ and $ \hat{X}_2 $ let $ \hat{X}'_1 $ and
$ \hat{X}'_2 $ 
denote their respective randomized versions according to the above
randomization.  Next, define $ \hat{j}^{'*}(z^n) $ to be
\[
\hat{j}^{'*}(z^n) = \arg \min_{j \in \{1,2\}} \hat{\overline{L}}_{\hat{X}'_j}(z^n)
\]
with $ \hat{\overline{L}}_{\hat{X}'_j} $
in~(\ref{eq:loss_estimate_randomized}) above. 
The estimation minimizing randomized denoiser is then defined as
\begin{equation}
\hat{X}^n_{RU}(z^n) = \hat{X}'_{\hat{j}^{'*}(z^n)}(z^n) =
\hat{X}_{\hat{j}^{'*}(z^n)}(z^n\oplus W^n).
\label{eq:randunivden}
\end{equation}
This denoiser thus determines the denoiser whose randomized version
yields the smallest estimated expected loss computed according
to~(\ref{eq:loss_estimate_randomized}) and denoises using the
randomized version of the selected denoiser.  We then have the
following. 
\begin{Lemma}
If for all $ \epsilon > 0 $, $ \hat{\overline{L}}_{\hat{X}'_j} $ satisfies
\begin{equation}
\limsup_{n\rightarrow \infty} 
\max_{x^n}\max_{j\in\{1,2\}} Pr(|\hat{\overline{L}}_{\hat{X}'_j}(Z^n) - 
\overline{L}_{\hat{X}'_j}(x^n,Z^n)| \geq \epsilon)
 = 0
\label{eq:conc1rand}
\end{equation}
then $ \hat{X}_{RU} $ satisfies
\begin{multline}
\limsup_{n\rightarrow \infty} \max_{x^n} E(\loss{\hat{X}_{RU}}(x^n,Z^n)) \\
- \min\{E(\loss{\hat{X}'_1}(x^n,Z^n)), 
E(\loss{\hat{X}'_2}(x^n,Z^n))\} = 0,
\label{eq:univdefrand}
\end{multline}
where the expectations are with respect to the channel output $ Z^n $ {\em and}
the randomization $ W^n $.
\label{lem:conc=univrand}
\end{Lemma}
\noindent{\bf Proof.}
Let $ j^{'*} $ denote
\[
j^{'*}(x^n,z^n){=}  \arg \min_{j \in \{1,2\}}
\overline{L}_{\hat{X}'_j}(x^n,z^n).
\]
Suppose for $ x^n $ and $ z^n $,
$ |\hat{\overline{L}}_{\hat{X}'_j}(z^n) {-} 
\overline{L}_{\hat{X}'_j}(x^n,z^n)| {\leq} \epsilon $ for $ j \in \{1,2\} $.
We then have
\begin{align*}
&\overline{L}_{\hat{X}'_{\hat{j}^{'*}}}(x^n,z^n)  - 
\overline{L}_{\hat{X}'_{j^{'*}}}(x^n,z^n) \\
&= \overline{L}_{\hat{X}'_{\hat{j}^{'*}}}(x^n,z^n){-}
 \hat{\overline{L}}_{\hat{X}'_{\hat{j}^{'*}}}(z^n) {+}
\hat{\overline{L}}_{\hat{X}'_{\hat{j}^{'*}}}(z^n) {-}
\overline{L}_{\hat{X}'_{j^{'*}}}(x^n,z^n) \\
&\leq \overline{L}_{\hat{X}'_{\hat{j}^{'*}}}(x^n,z^n) {-}
 \hat{\overline{L}}_{\hat{X}'_{\hat{j}^{'*}}}(z^n) {+}
\hat{\overline{L}}_{\hat{X}'_{j^{'*}}}(z^n) {-}
\overline{L}_{\hat{X}'_{j^{'*}}}(x^n,z^n) \\
& \leq 2\epsilon,
\end{align*}
implying, via a union bound, that
\begin{multline}
Pr(
\overline{L}_{\hat{X}'_{\hat{j}^{'*}}}(x^n,Z^n) - 
\overline{L}_{\hat{X}'_{j^{'*}}}(x^n,Z^n) \geq 2\epsilon) \\
\leq
\sum_{j=1}^2 
Pr(|\hat{\overline{L}}_{\hat{X}'_j}(Z^n) - \overline{L}_{\hat{X}'_j}(x^n,Z^n)| 
\geq \epsilon).
\label{eq:probineq2}
\end{multline}
Noting that $ \hat{X}'_{\hat{j}^{'*}} = 
\hat{X}_{RU} $,
it follows that,
for all $ \epsilon > 0 $,
\begin{align}
&\max_{x^n} E(\loss{\hat{X}_{RU}}(x^n,Z^n)) \nonumber \\
&\; \quad\quad - \min\{E(\loss{\hat{X}'_1}(x^n,Z^n)),
E(\loss{\hat{X}'_2}(x^n,Z^n))\} \nonumber \\
&\; \leq 
\max_{x^n} E(\overline{L}_{\hat{X}'_{\hat{j}^{'*}}}(x^n,Z^n) 
- \overline{L}_{\hat{X}_{j^{'*}}}(x^n,Z^n)) \label{eq:overlineequivstep} \\
&\; \leq
2\epsilon {+} \Lambda_{\max}\max_{x^n}
Pr(
\overline{L}_{\hat{X}'_{\hat{j}^{'*}}}(x^n,Z^n) {-}
\overline{L}_{\hat{X}_{j^{'*}}}(x^n,Z^n) \geq 2\epsilon) \label{eq:foralleps2}
\end{align}
where $ \Lambda_{\max} $ denotes the maximum loss and
(\ref{eq:overlineequivstep}) follows from
$
E(\loss{\hat{X}_{RU}}(x^n,Z^n)) {=} 
E(\overline{L}_{\hat{X}'_{\hat{j}^{'*}}}(x^n,Z^n))
$
and
\begin{align*}
&\min\{E(\loss{\hat{X}'_1}(x^n,Z^n)),
E(\loss{\hat{X}'_2}(x^n,Z^n))\}  \\
& =
\min\{E(\overline{L}_{\hat{X}'_1}(x^n,Z^n)),
E(\overline{L}_{\hat{X}'_2}(x^n,Z^n))\} \\
& \geq
E(\min\{\overline{L}_{\hat{X}'_1}(x^n,Z^n),
\overline{L}_{\hat{X}'_2}(x^n,Z^n)\}) {=} 
E(\overline{L}_{\hat{X}'_{j^{'*}}}(x^n,Z^n)).
\end{align*}
The lemma now follows from~(\ref{eq:probineq2}),
(\ref{eq:conc1rand}), and the fact that~(\ref{eq:foralleps2}) holds for all
$ \epsilon > 0 $.
$ \qed $

 This lemma shows that the loss 
estimation minimizing randomized denoiser exhibits the same asymptotic
expected performance as the  
best of two randomized denoisers, and if the expected performance of each such
randomized denoiser were, in turn, close to the
expected performance of the corresponding original denoiser, the
estimation minimizing randomized denoiser
would solve our original problem.
The proof of 
Lemma~\ref{lem:randden} is presented in the next section, while the
latter property is contained in the following.
\begin{Lemma}
For a BSC-$ \delta $, Hamming loss, and any denoiser $ \hat{X} $, if $ W^n $ is
i.i.d.\ Bernoulli-$q_n$ with $ q_n = n^{-\nu} $ for $ \nu > 1/2 $,
\[
\max_{x^n}|E(L(x^n,\hat{X}(Z^n{\oplus} W^n))){-}E(L(x^n,\hat{X}(Z^n)))| {=} o(1)
\]
where the first expectation is with respect to the channel and the
randomization.
\label{lem:closeexp}
\end{Lemma}

The proof of the lemma appears below.  It involves
showing that the $ L_1 $ distance between the distributions of the
random variables $ Z^n $ and $ Z^n \oplus W^n $ vanishes uniformly for
all input sequences $ x^n $.
\if{false}
\begin{Remark}
The lemma can be shown to hold for $ q_n = o(n^{-1/2}) $, but it
is a bit easier to write the proof if we concretely assume $ q_n =
n^{-\nu} $ for some $ \nu > 1/2 $. 
\end{Remark}
\fi
Thus, we have the following.
\begin{Theorem}
For $ \nu $ satisfying $ 1/2 < \nu < 1 $, the loss estimation
minimizing randomized denoiser $ \hat{X}_{RU} $ given
by~(\ref{eq:randunivden}),
with $ W^n $ i.i.d.\ Bernoulli-$n^{-\nu} $,
satisfies
\begin{multline*}
\limsup_{n\rightarrow \infty} \max_{x^n} E(\loss{\hat{X}_{RU}}(x^n,Z^n)) \\
- \min\{E(\loss{\hat{X}_1}(x^n,Z^n)), 
E(\loss{\hat{X}_2}(x^n,Z^n))\} = 0.
\end{multline*}
\label{thm:main}
\end{Theorem}

\noindent{\bf Proof of Lemma~\ref{lem:closeexp}:}
We start by noting that for any $ x^n $
\begin{align}
|E(L(x^n,&\hat{X}(Z^n\oplus W^n)))-E(L(x^n,\hat{X}(Z^n)))| \nonumber \\
& = |E(L(x^n,\hat{X}(\tilde{Z}^n)))-E(L(x^n,\hat{X}(Z^n)))| \nonumber \\
&\leq \sum_{z^n}|P_{\tilde{Z}^n}(z^n)-P_{Z^n}(z^n)| \label{eq:l1dist}
\end{align}
where $ \tilde{Z}^n = Z^n\oplus W^n $, and
in the last step $ P_{\tilde{Z}^n}(z^n) $ and $ P_{Z^n}(z^n) $ are the
respective probabilities of
$ \tilde{Z}^n = z^n $
and $ Z^n = z^n $ for the channel input sequence $x^n $.  It follows
from the properties of the channel that 
\begin{equation}
Pr(Z_i = 1) = \left\{\begin{array}{ll}
\delta & \mbox{if } x_i = 0 \\
\overline{\delta} & \mbox{if } x_i = 1 .
\end{array}
\right. 
\end{equation}
Letting
\[
v_n = \delta\overline{q_n}+\overline{\delta}q_n,
\]
it further follows from the channel and properties of $ W^n $
that $ \tilde{Z}^n $
are independent Bernoulli random variables with
\begin{equation}
Pr(\tilde{Z}_i = 1) = \left\{\begin{array}{ll}
v_n & \mbox{if } x_i = 0 \\
\overline{v_n}  & \mbox{if } x_i = 1 .
\end{array}
\right. 
\end{equation}

Notice that~(\ref{eq:l1dist}) is invariant to a permutation of the
underlying $ x^n $, so for notational convenience we shall assume that
$ x^m = 0 $ and 
$ x_{m+1}^n = 1 $, for some value of $ m $.
Define
\[
 A = \{z^n: |n_1(z^m) + n_0(z^{n}_{m+1})-\delta n| \leq n^{1/4+\nu/2}
 \},
\]
where for any binary sequence $ y^k $, $ n_1(y^k) $ and $ n_0(y^k) $
respectively denote the number of $ 1 $'s and $ 0 $'s in $ y^k $.
Since $ \nu > 1/2 $, the fact that $ n_1(Z^m)+n_0(Z^n_{m+1}) $ and
$ n_1(\tilde{Z}^m) + n_0(\tilde{Z}^n) $ respectively have the
same distributions as the sum of $ n $ i.i.d.\ Bernoulli-$ \delta $ and
$ n $ i.i.d.\ Bernoulli-$ v_n $
random variables along with 
standard results (e.g., Hoeffding's inequality)
imply that
\begin{equation}
 Pr(Z^n \in A^c) = o(1) \mbox{ and } Pr(\tilde{Z}^n \in A^c) = o(1).
\label{eq:Acompo1}
\end{equation}
In the case of the latter, note
that $ n v_n = n\delta + O(n^{1-\nu}) $ so that the deviation from the
mean implied by $ \tilde{Z}^n \in A^c $ is still $ O(n^{1/4+\nu/2}) $
(i.e., $ n^{1-\nu} = o(n^{1/4+\nu/2}) $ for $ \nu > 1/2 $).

Additionally, for $ z^n \in A $ we have
\begin{align}
\log \frac{P_{\tilde{Z}^n}(z^n)}{P_{Z^n}(z^n)} &= 
(n_1(z^m)+n_0(z_{m+1}^n))\log \left(\frac{v_n}{\delta}\right) \nonumber \\
& \quad
+(n_0(z^m)+n_1(z_{m+1}^n))\log
\left(\frac{\overline{v_n}}{\overline{\delta}}\right) \nonumber \\
&= (n\delta {+} d)\left(\frac{v_n{-}\delta}{\delta}{-}
\frac{(v_n{-}\delta)^2}{2\delta^2(1{+}\xi)^2}\right) \nonumber \\
& \quad +
(n\overline{\delta}{-}d)\left(-\frac{v_n{-}\delta}{\overline{\delta}}{-} 
\frac{(v_n{-}\delta)^2}{2\overline{\delta}^2(1+\xi')^2}\right) \label{eq:taylor}
\\ 
&=
d(v_n{-}\delta)\left(\frac{1-2\delta}{\delta\overline{\delta}}\right)
\nonumber \\
& \quad - (v_n{-}\delta)^2\left(\frac{n\delta {+} d}{2\delta^2(1{+}\xi)^2} {+}
\frac{n\overline{\delta}{-}d} {2\overline{\delta}^2(1{+}\xi')^2}
\right) \nonumber \\ 
& = o(1) \label{eq:laststepo1}
\end{align}
where~(\ref{eq:taylor}) follows by Taylor's approximation of $
\log(1+x) $ with $ d
\stackrel{\triangle}{=} $ $ n_1(z^m)+n_0(z_{m+1}^n) - n\delta $ and $
|\xi| \leq |(v_n{-}\delta)/\delta| $, $
|\xi'| \leq |(v_n{-}\delta)/\overline{\delta}| $
and~(\ref{eq:laststepo1}) follows since $ \nu > 1/2 $, which implies 
 $ |d(v_n{-}\delta)| = $ $ O(n^{1/4+\nu/2-\nu}) = $ $
O(n^{1/4-\nu/2}) = o(1) $ 
and $
n(v_n-\delta)^2 = O(n^{1-2\nu}) $ $ = o(1) $.

Applying these facts to~(\ref{eq:l1dist}), we obtain
\begin{align}
\sum_{z^n}|P_{\tilde{Z}^n}(z^n) & {-}P_{Z^n}(z^n)| \nonumber \\
&= o(1) + \sum_{z^n\in A}|P_{\tilde{Z}^n}(z^n)-P_{Z^n}(z^n)| \label{eq:probbnd}\\
&= o(1) + \sum_{z^n\in
  A}P_{Z^n}(z^n)\left|\frac{P_{\tilde{Z}^n}(z^n)}{P_{Z^n}(z^n)}
- 1\right| \nonumber \\
&= o(1) \label{eq:ratiobnd}
\end{align}
where~(\ref{eq:probbnd}) follows from~(\ref{eq:Acompo1})
and~(\ref{eq:ratiobnd}) follows from~(\ref{eq:laststepo1}), which is
uniformly vanishing for $ z^n \in A $, and the fact that $ e^{x} $ is
continuous.~$\qed$ 

\vspace{-.5cm}
\section{Proof of Lemma~\ref{lem:randden}}
We begin by defining, for any $ f:\{0,1\}^n \rightarrow \reals $,
the $ x^n $-dependent total influence (terminology inspired by a
related quantity in~\cite{KKL88}) of $ f $ as  
\[
I(f) = \sum_{j=1}^n E(|f(Z^n)-f(Z^{j-1},\tilde{Z}_j,Z_{j+1}^n)|)
\]
where $ (\tilde{Z}^n,Z^n)  $ constitute an i.i.d.\ pair of random
variables with $ Z^n $ distributed according to the channel
with input $ x^n $ (hence the dependence on $ x^n $).
The proof of Lemma~\ref{lem:randden} hinges on the following result.
\begin{Proposition}
For all $ 0 < \nu < 1 $ and $
\hat{X}'(z^n) $ defined as 
in~(\ref{eq:randdendef}),
\begin{equation}
\max_{x^n,i} I(\overline{\hat{X}'}(\cdot)[i]) = o(n),
\label{eq:infon}
\end{equation}
where 
$
\overline{\hat{X}'}(z^n)[i] {=} E_{W^n}(\hat{X}'(z^n)[i]) {=} 
E_{W^n}(\hat{X}(z^n\oplus W^n)[i]),
$
with the expectation taken with respect to $ W^n $.
\label{prop:infon}
\end{Proposition}

The proof, which follows, involves showing that
$ \max_{f}\max_{z^n} 
\sum_{j=1}^n|\overline{f}(z^n)-\overline{f}(z^n\oplus \blde_j)| = o(n), $
where
the outer maximization is over all 
functions $ f:\{0,1\}^n \rightarrow [0,1] $, with $
\overline{f}(z^n) \stackrel{\triangle}{=} E_{W^n}(f(z^n\oplus W^n)). $
This, in turn, is reduced to proving that the $ L_1 $ distance between
two related distributions vanishes.
\noindent{\bf Proof:}
For a function $ f:\{0,1\}^n \rightarrow [0,1] $, let $
\overline{f}(z^n) $ denote 
\[
\overline{f}(z^n) = E_{W^n}(f(z^n\oplus W^n)).  
\]
Also, let $ e_j $ denote the ``indicator'' sequence (or vector) with $
e_j[t] = 0 $ if 
$ t \neq j $ and $ e_j[j] = 1 $.
We will prove the proposition by showing that
\begin{equation}
\max_{f}\max_{z^n} \sum_{j=1}^n|\overline{f}(z^n)-\overline{f}(z^n\oplus e_j)| =
o(n),
\label{eq:mainstep}
\end{equation}
where the maximization over $ f $ is over all functions $
f:\{0,1\}^n \rightarrow [0,1] $.\footnote{A simple
  example (e.g., $ f(z^n) = z_1 $) shows that $ |\overline{f}(z^n)-\overline{f}(z^n\oplus
  e_j)| $ can be $ 
  \Omega(1) $ for any fixed $ j $, but it turns out this can't
  occur for too many $ j $'s simultaneously for any underlying $ f $.}

To see why~(\ref{eq:mainstep}) implies~(\ref{eq:infon}), note that
\begin{equation}
 \max_{\tilde{z}^n} \sum_{j=1}^n|\overline{f}(z^n)-\overline{f}(z^{j-1},\tilde{z}_j,z_{j+1}^n)| =
\sum_{j=1}^n|\overline{f}(z^n)-\overline{f}(z^n\oplus e_j)|
\label{eq:maxstepreason}
\end{equation}
and therefore,
\begin{align}
& I(\overline{\hat{X}'}(\cdot)[i]) \nonumber \\
& = 
E\Big(E\Big(\sum_{j=1}^n|\overline{\hat{X}'}(Z^n)[i]-\overline{\hat{X}'}(Z^{j-1},
\tilde{Z}_j,Z_{j+1}^n)[i]|\Big| Z^n\Big)\Big)  \nonumber\\ 
&\leq 
E\Big(\max_{\tilde{z}^n}\sum_{j=1}^n|\overline{\hat{X}'}(Z^n)[i]-
\overline{\hat{X}'}(Z^{j-1},\tilde{z}_j,Z_{j+1}^n)[i]|\Big)
\nonumber \\    
&=
E\Big(\sum_{j=1}^n|\overline{\hat{X}'}(Z^n)[i]-
\overline{\hat{X}'}(Z^n\oplus 
e_j)[i]|\Big)
\label{eq:maxstep}\\  
&\leq
\max_{f}\max_{z^n} \sum_{j=1}^n|\overline{f}(z^n)-\overline{f}(z^n\oplus e_j)| \label{eq:mainstepjust}
\end{align}
where~(\ref{eq:maxstep}) follows from~(\ref{eq:maxstepreason}).

We begin the proof of~(\ref{eq:mainstep}) with the observation that
\begin{align}
& \max_{f}\max_{z^n}
  \sum_{j=1}^n|\overline{f}(z^n)-\overline{f}(z^n\oplus e_j)| 
\nonumber \\
&= 
\max_{f}\sum_{j=1}^n|\overline{f}(0^n)-\overline{f}(e_j)| \nonumber \\
& = 
\max_{f}\max_{s^n \in \{-1,+1\}^n}
\sum_{j=1}^ns_j(\overline{f}(0^n)-\overline{f}(e_j)) \nonumber \\
&= \max_{s^n \in \{-1,+1\}^n} \max_f
    \sum_{j=1}^ns_j(E(f(W^n))-E(f(W^n\oplus e_j))) \label{eq:maxovers}
\end{align}
where the expectations are with respect to $ W^n $.  

Next, we note that for any  $ s^n $, $ f $, and permutation $ \sigma $
of $ (1,2,\ldots,n) $
\begin{multline}
\sum_{j=1}^ns_{\sigma(j)}(E(f(W^n))-E(f(W^n\oplus e_j))) \\ =
\sum_{j=1}^ns_{j}(E(f\circ\sigma^{-1}(W^n))-E(f\circ\sigma^{-1}(W^n\oplus
e_j)))
\label{eq:perminv}
\end{multline}
where 
\[ f\circ\sigma^{-1}(z_1,\ldots,z_n) =
f(z_{\sigma^{-1}(1)},\ldots,z_{\sigma^{-1}(n)})
\]
and $ \sigma^{-1} $ is the inverse permutation of $ \sigma $.
This can be seen as follows:
\begin{align}
\sum_{j=1}^ns_{j}&(E(f\circ\sigma^{-1}(W^n))-E(f\circ\sigma^{-1}(W^n\oplus
e_j))) \nonumber \\
 &= \sum_{j=1}^ns_{j}(E(f(W^n))-E(f(W^n\oplus
e_{\sigma^{-1}(j)}))) \label{eq:perminvw} \\
&= \sum_{j=1}^ns_{\sigma(j)}(E(f(W^n))-E(f(W^n\oplus
e_j))), \nonumber
\end{align}
where~(\ref{eq:perminvw}) follows from the fact that the distribution
of $ W^n $ is permutation invariant.  

Relation~(\ref{eq:perminv})
implies that the maximization over $ s^n $ in~(\ref{eq:maxovers}) 
can be restricted to $ s^n $ for which $ s_j = 1 $ for $ j \leq m $
and $ s_j = -1 $ for $ j > m $, for some $ m \in \{0,1,\ldots,n\}$.
Given such an $ m $, the maximization over $ f $
in~(\ref{eq:maxovers}) can be expressed as
\begin{align}
\max_f \Big[
    \sum_{j=1}^m & E(f(W^n))-E(f(W^n\oplus e_j)) \nonumber \\ & \quad + 
    \sum_{j=m+1}^n E(f(W^n\oplus e_j)) - E(f(W^n)) \Big] \nonumber \\
&= \max_f n(E(f(W_1^n)) - E(f(W_2^n))) \nonumber \\
&\leq \frac{n}{2} \sum_{w^n}|p_1(w^n)-p_2(w^n)| \label{eq:l1bndws} 
\end{align}
where $ W_1^n $ and $ W_2^n $ are random sequences with respective
probability distributions $ p_1(\cdot) $ and $ p_2(\cdot) $, and where
$ W_1^n = W^n $ with probability $ m/n $ and 
$ W_1^n = W^n\oplus e_j $ with probability $ 1/n $ for $ j \in
\{m{+}1,\ldots, n \} $,  $ W_2^n = W^n $ with probability $ 1-m/n $
and $ W_2^n = W^n\oplus e_j $ with probability $ 1/n $ for $ j \in
\{1,\ldots, m\} $.
The last step~(\ref{eq:l1bndws}) follows since the range of $ f $ is
in $ [0,1] $ (any bounded range could be accounted for with a
suitable constant factor).

Recalling the definition of $ W^n$, we have that $ p(w^n) =
q_n^{n_1(w^n)}\overline{q_n}^{n-n_1(w^n)} $.   
It then follows from the above definitions of $ W_1 $ and $ W_2 $ that
\begin{align}
& p_1(w^n) \nonumber \\
&= \frac{m}{n}p(w^n) + \frac{1}{n}\sum_{j=m+1}^n p(w^n\oplus e_j)
\nonumber \\
&= p(w^n)\Bigg[\frac{m}{n}{+} \frac{1}{n}\Bigg(
n_1(w_{m{+}1}^n)\frac{\overline{q_n}}{q_n} {+} 
(n{-}m{-}n_1(w_{m{+}1}^n))\frac{q_n}{\overline{q_n}} \Bigg) \Bigg]  
\label{eq:w1prob}
\end{align}
and
\begin{align}
& p_2(w^n) \nonumber \\
&= \frac{n-m}{n}p(w^n) + \frac{1}{n}\sum_{j=1}^m p(w^n\oplus e_j)
\nonumber \\
&= p(w^n)\Bigg[\frac{n-m}{n}+ \frac{1}{n}\Bigg(
n_1(w^m)\frac{\overline{q_n}}{q_n} 
+
(m-n_1(w^m))\frac{q_n}{\overline{q_n}} \Bigg) \Bigg].
\label{eq:w2prob}
\end{align}
These imply the obvious bounds
\begin{multline}
p(w^n)\left[\frac{m}{n}+ \frac{n_1(w_{m+1}^n)}{n}\frac{\overline{q_n}}{q_n}
\right] \leq p_1(w^n) \\ \leq
p(w^n)\left[\frac{m}{n}+
  \frac{n_1(w_{m+1}^n)}{n}\frac{\overline{q_n}}{q_n}+\frac{q_n}{\overline{q_n}}
  \right]   
\label{eq:w1probbnds}
\end{multline}
and
\begin{multline}
p(w^n)\left[\frac{n-m}{n}+ \frac{n_1(w^m)}{n}\frac{\overline{q_n}}{q_n}
\right] \leq p_2(w^n)  \\ \leq
p(w^n)\left[\frac{n-m}{n}+
  \frac{n_1(w^m)}{n}\frac{\overline{q_n}}{q_n}+\frac{q_n}{\overline{q_n}}
  \right], 
\label{eq:w2probbnds}
\end{multline}
which, in turn, imply
\begin{align}
&\sum_{w^n}|p_1(w^n)-p_2(w^n)| \nonumber \\
&\leq
\frac{q_n}{\overline{q_n}} {+} E\Bigg[\Bigg|\frac{m}{n}{+}
  \frac{n_1(W_{m{+}1}^n)}{n}\frac{\overline{q_n}}{q_n} 
{-}\frac{n{-}m}{n}
  {-}\frac{n_1(W^m)}{n}\frac{\overline{q_n}}{q_n}
  \Bigg|\Bigg] \nonumber \\
&= \frac{q_n}{\overline{q_n}} {+} E\Bigg[\Bigg|\frac{m}{n} {+}
  \frac{(n_1(W_{m{+}1}^n){-}(n{-}m)q_n{+}(n{-}m)q_n)}{n}\frac{\overline{q_n}}{q_n} 
\nonumber \\
& \quad\quad\quad\quad\quad {-}\frac{n{-}m}{n}
  {-}\frac{(n_1(W^m){-}mq_n{+}mq_n)}{n}\frac{\overline{q_n}}{q_n}
  \Bigg|\Bigg] \nonumber \\
&= \frac{q_n}{\overline{q_n}} {+} E\Bigg[\Bigg|
  \frac{(n_1(W_{m{+}1}^n){-}(n{-}m)q_n)}{n}\frac{\overline{q_n}}{q_n}
\nonumber \\
& \quad\quad\quad\quad\quad 
  {-}\frac{(n_1(W^m){-}mq_n)}{n}\frac{\overline{q_n}}{q_n} {+} 
 q_n\frac{2m{-}n}{n}  \Bigg|\Bigg] \nonumber \\
 &\leq \frac{2q_n}{\overline{q_n}} {+} \frac{1}{nq_n}(
E[|n_1(W_{m{+}1}^n){-}q_n(n{-}m)|] \nonumber \\
& \mbox{\hspace{1in}} {+} E[|n_1(W^m){-}q_nm|]),
\label{eq:devstep}
\end{align}
where the expectations are with respect to $ W^n $.

We will bound the expectations in~(\ref{eq:devstep}) using the concentration
inequality~\cite[Theorem 2.3]{mcdiarmid}
\begin{equation}
P\Big(|n_1(W^k) - kq_n| \ge \epsilon \Big) \le
2\exp\left( - \frac { \epsilon^2}{2k q_n \left(1+ \epsilon/(3kq_n)
  \right)}\right), 
\label{eq:mcdiarmid}
\end{equation}
which is applicable since $ W^n $ is i.i.d.\ with $ W_j \in \{0,1\} $.
Using the well known integration-by-parts formula for the expectation of a
non-negative random variable, we have
\begin{align}
E&[|n_1(W^k){-}q_nk|] \nonumber \\
&= \int_{0}^{\infty} P\Big(|n_1(W^k) {-}
kq_n| \ge \epsilon \Big) d\epsilon \nonumber \\
& \leq \int_{0}^{\infty}
2\exp\left( {-} \frac { \epsilon^2}{2k q_n \left(1{+} \epsilon/(3kq_n)
  \right)}\right) d\epsilon \nonumber \\
& \leq \int_{0}^{kq_n}
2\exp\left( {-} \frac { 3\epsilon^2}{8k q_n}\right) d\epsilon 
{+} \int_{kq_n}^{\infty}
2\exp\left( {-} \frac { 3\epsilon}{8}\right) d\epsilon  \nonumber \\
& \leq \sqrt{\frac{8\pi kq_n}{3}} 
{+} \frac{16}{3}\exp\left({-}\frac{3kq_n}{8}\right).  \nonumber 
\end{align}
Applying this in~(\ref{eq:devstep}) with $ k = n-m $ and $ k = m $,
respectively, yields
\begin{align}
&\sum_{w^n}|p_1(w^n){-}p_2(w^n)| \nonumber \\
&\leq
\frac{2q_n}{\overline{q_n}} {+} \frac{1}{nq_n}\sqrt{\frac{8\pi (n{-}m)q_n}{3}} 
{+} \frac{1}{nq_n}\frac{16}{3}\exp\left({-}\frac{3(n{-}m)q_n}{8}\right)
\nonumber \\
& \quad\quad {+} \frac{1}{q_nn}\sqrt{\frac{8\pi m q_n}{3}} 
{+} \frac{1}{nq_n}\frac{16}{3}\exp\left({-}\frac{3mq_n}{8}\right) \nonumber \\
&= O(q_n) {+} O((nq_n)^{-1/2}) {+} O((nq_n)^{-1}) \nonumber \\
&= o(1), \nonumber
\end{align}
uniformly in $ m $,
where the last step follows from our assumption that $ q_n = n^{-\nu}
$ for $ 0 < \nu < 1 $.   Incorporating this bound
into~(\ref{eq:l1bndws}), and then into~(\ref{eq:maxovers}), combined
with the observation~(\ref{eq:perminv}),
establishes~(\ref{eq:mainstep}) via~(\ref{eq:mainstepjust}),
completing the proof.~$\qed$

\noindent{\bf Proof of Lemma~\ref{lem:randden}:}
The proof is similar to that of Proposition~\ref{prop:functionoffew},
except the correlations appearing in~(\ref{eq:zerocorra}) are
handled using Proposition~\ref{prop:infon}.
Define
\begin{equation}
\overline{\Delta}_i(z^n) \stackrel{\triangle}{=}
\overline{\tilde{\Lambda}}_{i,\re'}\Paren{z^n} -
E_{W^n}\Lambda\Paren{x_i,\re(z^n\oplus W^n)[i]}
\end{equation}
with~$\overline{\tilde{\Lambda}}_{i,\re'}$ as
in~(\ref{eq:overlinetildelambda}).   We claim that
\begin{equation}
\max_{x^n,i} I(\overline{\Delta}_i(\cdot)) = o(n).
\label{eq:deltainfon}
\end{equation}
To see this, note that for the binary/Hamming loss case, $
E_{W^n}\Lambda(x,\hat{X}(z^n\oplus W^n)[i]) =
  \overline{\hat{X}'}(z^n)[i] $ if $ x = 0 $ and  
$ 1 - \overline{\hat{X}'}(z^n)[i] $ if $ x = 1 $.  
It is then
  immediate from the definitions that $ \overline{\Delta}_i $ can be
  expressed as 
\begin{multline}
\overline{\Delta}_{i}(z^n) = c_1(z_i) +
c_2(z_i)\overline{\hat{X}'}(z^{i-1},0,z_{i+1}^n)[i] \\ + 
c_3(z_i)\overline{\hat{X}'}(z^{i-1},1,z_{i+1}^n)[i] + 
c_4(x_i) + c_5(x_i)\overline{\hat{X}'}(z^n)[i] 
\label{eq:overlinedeltaexpr}
\end{multline}
for $ z_i $ and $ x_i $ dependent quantities $ c_1, \ldots, c_5 $.
Thus, we have
\begin{align}
&I(\overline{\Delta}_i(\cdot)) \nonumber \\ &=
\sum_{j=1}^nE(|\overline{\Delta}_{i}(Z^n) {-}
\overline{\Delta}_{i}(Z^{j{-}1},\tilde{Z}_j,Z_{j{+}1}^n)|) \nonumber \\
&\leq d_1
{+} d_2\sum_{j\neq i}E\big(|\overline{\hat{X}'}(Z^n)[i]{-}
\overline{\hat{X}'}(Z^{j{-}1},\tilde{Z}_j,Z_{j{+}1}^n)[i]|\big|Z_i =
0\big) \nonumber \\
& \quad
{+} d_3\sum_{j\neq i}E\big(|\overline{\hat{X}'}(Z^n)[i]{-}
\overline{\hat{X}'}(Z^{j{-}1},\tilde{Z}_j,Z_{j{+}1}^n)[i]|\big|Z_i =
1\big) \nonumber \\
& \quad {+} d_4\sum_{j\neq i}E\big(|\overline{\hat{X}'}(Z^n)[i]{-}
\overline{\hat{X}'}(Z^{j{-}1},\tilde{Z}_j,Z_{j{+}1}^n)[i]|\big) 
\label{eq:IDeltastep1} \\
&\leq d_1 {+} d_5I(\overline{\hat{X'}}(\cdot)[i]),
\label{eq:IDeltalaststep}
\end{align}
where $ d_1,\ldots,d_5 $ are bounded $ x_i $ dependent quantities and
where~(\ref{eq:IDeltastep1}) follows from~(\ref{eq:overlinedeltaexpr})
and the triangle inequality.  The claim~(\ref{eq:deltainfon}) follows
from~(\ref{eq:IDeltalaststep}) and Proposition~\ref{prop:infon} since
$ d_1 $ and $ d_5 $ can be bounded 
uniformly in $ x^n $ and $ i $.

Next, we note that for $ (Z^n,\tilde{Z}^n) $ an i.i.d.\ pair with $
Z^n $ distributed according to the channel (as in the definition of
total influence above), for all pairs $ (i,j) $,
\begin{align}
&E(\overline{\Delta}_i(Z^{j{-}1},\tilde{Z}_j,Z_{j{+}1}^n)
\overline{\Delta}_j(Z^n)) \nonumber \\
&=  
E\big(E\big(\overline{\Delta}_i(Z^{j{-}1},\tilde{Z}_j,Z_{j{+}1}^n)
\overline{\Delta}_j(Z^n)\big|  
Z^{j{-}1},\tilde{Z}_j,Z_{j{+}1}^n\big)\big) \nonumber \\
&= 
E\big(\overline{\Delta}_i(Z^{j{-}1},\tilde{Z}_j,Z_{j{+}1}^n)
E\big(\overline{\Delta}_j(Z^n)\big| 
Z^{j{-}1},\tilde{Z}_j,Z_{j{+}1}^n\big)\big) \nonumber \\
&= 0 \label{eq:overlinedeltauncorr},
\end{align}
where this last step follows from the conditional
unbiasedness~(\ref{eq:conditional_unbiased_rand}) and the
distribution of $ (Z^n,\tilde{Z}^n) $.

We then have
\begin{align}
&E\Big(\sum_{i=1}^n\overline{\Delta}_i(Z^n)\Big)^2 \nonumber \\ &= 
\sum_{i=1}^nE\Big(\sum_{j=1}^n\overline{\Delta}_i(Z^n)\overline{\Delta}_j(Z^n)
\Big) \nonumber \\
&= \sum_{i=1}^nE\Big(
\sum_{j=1}^n\overline{\Delta}_i(Z^j,\tilde{Z_j},Z_j^n)\overline{\Delta}_j(Z^n)
\Big) \nonumber \\
& \quad {+} \sum_{i=1}^nE\Big(
\sum_{j=1}^n(\overline{\Delta}_i(Z^n){-}
\overline{\Delta}_i(Z^j,\tilde{Z_j},Z_j^n))\overline{\Delta}_j(Z^n)  
\Big) \nonumber \\
&= \sum_{i=1}^nE\Big(
\sum_{j=1}^n(\overline{\Delta}_i(Z^n){-}
\overline{\Delta}_i(Z^j,\tilde{Z_j},Z_j^n))\overline{\Delta}_j(Z^n)  
\Big) \label{eq:uncorrstep} \\
&\leq c\sum_{i=1}^nE\Big(
\sum_{j=1}^n(|\overline{\Delta}_i(Z^n){-}
\overline{\Delta}_i(Z^j,\tilde{Z_j},Z_j^n)|)
\Big) \nonumber \\
&\leq c\sum_{i=1}^nI(\overline{\Delta}_i(\cdot)),
\label{eq:Idefstep} 
\end{align}
where~(\ref{eq:uncorrstep}) follows
from~(\ref{eq:overlinedeltauncorr}) and (\ref{eq:Idefstep}) from the fact
that $ \overline{\Delta}_i(z^n) $ can be bounded by a constant $ c $ 
for all $ i $, $
z^n $ and $ x^n $ and the
definition of total influence. The proof is completed by 
applying~(\ref{eq:deltainfon}).~$\qed$

\end{document}